\documentclass[12pt]{article}

\usepackage{amsmath}
\usepackage{amssymb}
\usepackage{epsfig}
\usepackage{graphicx}
\usepackage{cite}

\allowdisplaybreaks[1]

\makeatletter
\renewcommand{\fnum@table}{\textbf{\tablename~\thetable}}
\renewcommand{\fnum@figure}{\textbf{\figurename~\thefigure}}
\makeatother

\newcounter{myenumi}

\renewcommand{\themyenumi}{\roman{myenumi}}
{\end{list}}

\newlength{\myem}
\newcounter{mysubequation}[equation]

\makeatletter
\renewcommand{\section}{\@startsection{section}{1}{0em}{-\baselineskip}%
{\baselineskip}{\normalfont\large\bfseries}}
\renewcommand{\subsection}%
{\@startsection{subsection}{2}{0em}{-0.7\baselineskip}%
{0.7\baselineskip}{\normalfont\bfseries}}
\makeatother

\newcommand{\bi}{\begin{itemize}}
\newcommand{\ei}{\end{itemize}}

\newcommand{\be}{\begin{equation}}
\newcommand{\ee}{\end{equation}}
\newcommand{\bea}{\begin{eqnarray}}
\newcommand{\eea}{\end{eqnarray}}

\newcommand{\ldm}{\Delta m_{31}^2}
\newcommand{\sdm}{\Delta m_{21}^2}
\newcommand{\deltacp}{\delta_{\mathrm{CP}}}
\newcommand{\stheta}{\sin^2(2 \theta_{13})}
\newcommand{\sthetat}{\sin^2(2 \tilde{\theta}_{13})}
\newcommand{\CP}{\emph{CP}}

\newcommand{\eps}[0]{\epsilon}
\newcommand{\ket}[1]{\left|#1\right>}

\newcommand{\braket}[2]{\left<#1|#2\right>}
\newcommand{\obraket}[3]{\left<#1\left|#2\right|#3\right>}

\newcommand{\ie}{{\it i.e.}}

\newcommand{\eg}{{\it e.g.}}

\newcommand{\cf}{{\it cf.}}

\newcommand{\etc}{{\it etc.}}
\newcommand{\eq}{Eq.}
\newcommand{\eqs}{Eqs.}

\newcommand{\fig}{Fig.}
\newcommand{\Fig}{Fig.}

\newcommand{\Ref}{Ref.}
\newcommand{\Refs}{Refs.}
\newcommand{\Sec}{Sec.}

\newcommand{\App}{Appendix}

\newcommand{\Tab}{Table}

\newcommand{\equ}[1]{\eq~(\ref{equ:#1})}
\newcommand{\figu}[1]{\fig~\ref{fig:#1}}

\newcommand{\mtrx}[2]{\left(\begin{array}{#1} #2 \end{array}\right)}

\DeclareMathOperator{\diag}{diag}
\DeclareMathOperator{\tr}{tr}
\DeclareMathOperator{\sgn}{sgn}
\DeclareMathOperator{\real}{Re}
\DeclareMathOperator{\imag}{Im}

\begin{document}

\begin{titlepage}

\renewcommand{\thefootnote}{\alph{footnote}}

\renewcommand{\thefootnote}{\fnsymbol{footnote}}
\setcounter{footnote}{-1}

{\begin{center}
{\large\bf
Non-standard Hamiltonian effects on neutrino oscillations
} \end{center}}
\renewcommand{\thefootnote}{\alph{footnote}}

\vspace*{.8cm}
\vspace*{.3cm}
{\begin{center} {\large{\sc
 		    Mattias~Blennow\footnote[1]{\makebox[1.cm]{Email:}
                emb@kth.se},
                Tommy~Ohlsson\footnote[2]{\makebox[1.cm]{Email:}
                tommy@theophys.kth.se}, and
                Walter~Winter\footnote[3]{\makebox[1.cm]{Email:}
                winter@ias.edu}~
                }}
\end{center}}
\vspace*{0cm}
{\it
\begin{center}

\footnotemark[1]${}^,$\footnotemark[2]%
Department of Theoretical Physics, School of
Engineering Sciences, \\
Royal Institute of Technology (KTH) -- AlbaNova
University Center,\\
Roslagstullsbacken 21, 106~91~~Stockholm, Sweden

\footnotemark[3]%
       School of Natural Sciences, Institute for Advanced Study, \\
       Einstein Drive, Princeton, NJ 08540, USA

\end{center}}

\vspace*{1.5cm}

{\Large \bf
\begin{center} Abstract \end{center}}

We investigate non-standard Hamiltonian effects on neutrino
oscillations, which are effective additional contributions to the
vacuum or matter Hamiltonian. Since these effects can enter in either
flavor or mass basis, we develop an understanding of the difference
between these bases representing the underlying theoretical model. In
particular, the simplest of these effects are classified as 
``pure'' flavor or mass effects, where the appearance of such a ``pure''
effect can be quite plausible as a leading non-standard contribution
from theoretical models. Compared to earlier studies investigating
particular effects, we aim for a top-down classification of a possible
``new physics'' signature at future long-baseline neutrino oscillation
precision experiments.  We develop a general framework for such effects
with two neutrino flavors and discuss the extension to three neutrino
flavors, as well as we demonstrate the challenges for a neutrino
factory to distinguish the theoretical origin of these effects with a
numerical example.
We find how the precision measurement of neutrino oscillation parameters can be altered by non-standard effects alone (not including non-standard interactions in the creation and detection processes) and that the non-standard effects on Hamiltonian level can be distinguished from other non-standard effects (such as neutrino decoherence and decay) if we consider specific imprint of the effects on the energy spectra of several different oscillation channels at a neutrino factory.

\vspace*{.5cm}

\end{titlepage}

\newpage

\renewcommand{\thefootnote}{\arabic{footnote}}
\setcounter{footnote}{0}

\section{Introduction}

Neutrino physics has entered the era of precision measurements of the
fundamental neutrino parameters such as neutrino mass squared
differences and leptonic mixing parameters, and neutrino
oscillations are the most credible candidate for describing neutrino
flavor transitions. Nevertheless, there might be other sub-leading
mechanisms participating in the total description of neutrino flavor
transitions. Thus, in this paper, we will investigate such mechanisms
on a fundamental level, which will give rise to non-standard effects
on the ordinary framework of neutrino oscillations.

In a previous paper \cite{Blennow:2005yk}, we have studied
non-standard effects on probability level based on ``damping
signatures'', which were phenomenologically introduced in the neutrino
oscillation probabilities. However, in this paper, we will investigate
so-called non-standard Hamiltonian effects, which are effects on
Hamiltonian level rather than on probability level. Recently, three
different main categories of non-standard Hamiltonian effects have
been discussed in the literature. These categories are non-standard
interactions (NSI), flavor changing neutral currents (FCNC), and mass
varying neutrinos (MVN or MaVaNs). In addition, other effects which result in effective additions to the Hamiltonian have been studied,
such as from extra dimensions~\cite{Pas:2005rb}.
Below, we will shortly review the categories of effects which can be
studied using this framework.

In general, in many models, neutrino masses come together with NSI,
which means that the evolution of neutrinos passing through matter is
modified by non-standard potentials due to coherent forward-scattering
of NSI processes $\nu_\alpha + f \to \nu_\beta + f$, where
$\alpha,\beta = e,\mu,\tau$ and $f$ is a fermion in matter.\footnote{Note that, in general,
the production and detection vertices could also be modified. However, in this paper, we focus on the neutrino oscillation probabilities which, in the limit of ultrarelativistic neutrinos, decouple from the creation and detection processes.} The
effective NSI potentials are given by $V_{\rm NSI} = \sqrt{2} G_F N_d
\tilde\epsilon_{\alpha\beta}$, where $G_F$ is the Fermi coupling constant,
$N_d$ is the down quark number density, and $\tilde\epsilon_{\alpha\beta}$'s
are small parameters describing the NSI \cite{Valle:1987gv}. See, \eg,
\Ref~\cite{Valle:2003uv} for a recent review. Furthermore,
matter-enhanced neutrino oscillations in presence of $Z$-induced FCNC
have been studied in the literature
\cite{Bergmann:1997mr,Bergmann:1998rg,Bergmann:1999rz}. See also, \eg,
\Refs~\cite{Wolfenstein:1977ue,Krastev:1997cp} for some earlier
contributions. Especially, NSI and FCNC have been investigated in
several references for many different scenarios such as for solar~\cite{Bergmann:2000gp,Guzzo:2001mi,Friedland:2004pp,Miranda:2004nb},
atmospheric~\cite{Bergmann:1999pk,Fornengo:2001pm,Gonzalez-Garcia:2004wg,Friedland:2004ah,Friedland:2005vy},
supernova~\cite{Fogli:2002xj}, and other astrophysical neutrinos as
well as for \CP{} violation~\cite{Bekman:2002zk}, the LSND experiment~\cite{Bergmann:1998ft}, beam experiments~\cite{Ota:2002na}, and neutrino factories~\cite{Ota:2001pw,Gonzalez-Garcia:2001mp,Huber:2001zw,Huber:2001de,Huber:2002bi,Campanelli:2002cc}.

The idea of MVN was proposed by Fardon {\it et al.}~in
\Refs~\cite{Gu:2003er,Fardon:2003eh}. This idea is based on the dark energy of
the Universe being neutrinos which can act as a negative pressure
fluid and be the origin of cosmic acceleration. Furthermore, several
continuation works on MVN have been performed in the context of
scenarios for the Sun and the solar neutrino deficit
\cite{Barger:2005mn,Cirelli:2005sg}, but also in various other
contexts
\cite{Bi:2003yr,Hung:2003jb,Kaplan:2004dq,Gu:2004xx,Zurek:2004vd,Peccei:2004sz,Li:2004tq,Bi:2004ns,Horvat:2005ua,Afshordi:2005ym,Takahashi:2005kw,Fardon:2005wc,Brookfield:2005td}.
In addition, it should be mentioned that neutrinos with variable
masses have also been studied earlier than the idea of MVN
\cite{Kawasaki:1991gn,Stephenson:1996qj,Stephenson:1996cz,Sawyer:1998ac,Gu:2003er}.

While earlier studies have discussed individual theoretical
models and their effects on future neutrino oscillation experiments
(bottom-up), our approach will be the top-down. We start from
general assumptions to investigate the properties of non-standard
Hamiltonian effects, and later apply them to specific models and
discuss how to identify individual effects. The goal of this approach
is the classification of a possible ``new physics'' signature in future
long-baseline neutrino oscillation experiments. Although it is very
likely that such a signature will fit many different non-standard
models, it has hardly been discussed in the literature how to
distinguish (even qualitatively) different theoretical models which
could all describe this effect, and what the methods for that
identification could be.  For this purpose, we make rather unspecific
assumptions for the particular type of effect and rather assume that
the theoretical model will predict a leading effect which can be considered to be of a ``simple'' form in a specific
basis (``pure'' effect), which can be either flavor (or mass)
conserving or flavor (or mass) violating.

The paper is organized as follows: First, in
\Sec~\ref{sec:nonstandard_n}, we define non-standard Hamiltonian
effects as effective additional contributions to the vacuum
Hamiltonian similar to matter effects. The definition is performed for
$n$ neutrino flavors. Next, in \Sec~\ref{sec:nonstandard_2}, we
specialize our discussion to two neutrino flavors, where we derive the
effective neutrino parameters as well as the resonance conditions in
both flavor and mass bases including non-standard Hamiltonian
effects. We also discuss experimental strategies to test and
identify non-standard Hamiltonian effects at the example of
$\nu_e \leftrightarrow \nu_\mu$ flavor transitions.
Then, in \Sec~\ref{sec:threeflavtheory}, we study some aspects of the
generalization to three-flavor case, whereas in
\Sec~\ref{sec:numex}, we give a numerical example of how non-standard
Hamiltonian effects can affect a realistic experimental
setup and discuss how to tell non-standard Hamiltonian effects apart from damping effects. Finally, we summarize our results and present our conclusions in \Sec~\ref{sec:S&C}.

\section{Parameterization of non-standard Hamiltonian effects}
\label{sec:nonstandard_n}

In the standard neutrino oscillation framework with $n$ flavors, the
Hamiltonian in vacuum is given by
\begin{equation}
H_0 = \frac{1}{2E} U\diag(m_1^2, m_2^2, \ldots, m_n^2)U^\dagger \label{equ:hamvac}
\end{equation}
in flavor basis, where $E$ is the neutrino energy, $U$ is the leptonic
mixing matrix, and $m_i$ is the mass of the $i$th neutrino mass
eigenstate. Any Hermitian non-standard Hamiltonian effect will alter
this vacuum Hamiltonian into an effective Hamiltonian
\begin{equation}
H_{\rm eff} = H_0 + H', \label{equ:ns}
\end{equation}
where $H'$ is the effective addition to the vacuum Hamiltonian. We
note that this reminds of neutrino mixing and oscillations in matter
\cite{Wolfenstein:1977ue} with $H'$ given by a diagonal matrix with
the effective matter potentials on the diagonal, \ie,
\begin{equation}
H'=H_{\rm mat} = \diag(V,0,\ldots,0) - \frac{1}{\sqrt{2}} G_F N_n
\boldsymbol 1_n,
\end{equation}
where $V = \sqrt{2} G_F N_e$ is the ordinary matter potential, $G_F$
is the Fermi coupling constant, $N_e$ is the electron number density
(resulting from coherent forward-scattering of neutrinos), $N_n$ is
the nucleon number density, and $\boldsymbol 1_n$ is the $n \times n$
unit matrix\footnote{If sterile neutrinos are present, then
there is no interaction between the sterile neutrinos and the matter
through which they propagate. Thus, the $\boldsymbol 1_n$ term is
replaced by a projection operator on the active neutrino states.}. Just as the presence of matter affects the effective neutrino
mixing parameters, the effective neutrino mixing parameters will be
affected by any non-standard Hamiltonian effect. In the remainder of this text, we will treat the effective Hamiltonian
\begin{equation}
H_{\rm eff} = H_0 + H' + H_{\rm mat},
\end{equation}
\ie, we will treat the non-standard effects along with the matter effects. However, in \Sec~\ref{sec:threeflavtheory}, we treat only the part $H_0 + H'$ in order to obtain the parameters of the Hamiltonian to which the standard matter effects are then added. Since standard matter effects are generally taken into account, $H_0+H'$ will be mistaken for the vacuum Hamiltonian $H_0$ if the non-standard effects are not considered.

Since any part of the effective Hamiltonian that is proportional to
the $n\times n$ unit matrix only contributes with an overall phase to
the final neutrino state, it will not affect the neutrino oscillation
probabilities. This means that we may assume $H'$ to be traceless and
also that we may subtract $\tr(H)/n$ from the effective Hamiltonian to
make it traceless. Any traceless Hermitian $n\times n$ matrix $A$ may
be written as
\begin{equation}
A = \sum_{i = 1}^N c_i \lambda_i,
\label{equ:superposition}
\end{equation}
where the $c_i$'s are real numbers, the $\lambda_i$'s are the
generators of the $su(n)$ Lie algebra (\ie, $A$ is an element of the
Lie algebra), and $N = n^2-1$ is the number of generators. Hence,
clearly, any non-standard Hamiltonian effect $H'$ is parameterized by
the $n^2 - 1$ numbers $c_i$. In summary, we choose the coefficients of
the generators of the $su(n)$ Lie algebra to parameterize any
non-standard Hamiltonian effect.

Furthermore, in any basis (\eg, flavor or mass basis), we may introduce $su(n)$
generators $\lambda_i$ such that $n(n-1)/2$ generators are
off-diagonal with only two real non-zero entries, $n(n-1)/2$
generators are off-diagonal with only two imaginary non-zero entries,
and $n-1$ generators are diagonal with real entries. For example, in
the case of $n=2$, we have the Pauli matrices
\begin{equation}
\lambda_1 = \sigma_1 = \left(
\begin{array}{cc}
0 & 1 \\ 1 & 0
\end{array}
\right), \quad
\lambda_2 = \sigma_2 = \left(
\begin{array}{cc}
0 & -{\rm i} \\ {\rm i} & 0
\end{array}
\right), \quad
\lambda_3 = \sigma_3 = \left(
\begin{array}{cc}
1 & 0 \\ 0 & -1
\end{array}
\right).
\label{equ:pauli}
\end{equation}
We will denote the set of generators which are of the form $\lambda_i$
in flavor basis by $\rho_i$ and the set of generators which are of
this form in mass basis by $\tau_i$. Obviously, in flavor basis,
we have the relations
\begin{equation}
\rho_i = \lambda_i \quad {\rm and} \quad \tau_i = U \lambda_i U^\dagger,
\label{equ:sigma_tau}
\end{equation}
where, in the case of two neutrino flavors,
$$
U = \left( \begin{array}{cc} \cos(\theta) & \sin(\theta) \\
  -\sin(\theta) & \cos(\theta) \end{array} \right)
$$
is the two-flavor leptonic mixing matrix and $\theta$ is the
corresponding mixing angle (when treating the three-flavor case, we
will use the standard parameterization of the leptonic mixing with
three mixing angles $\theta_{12}$, $\theta_{23}$, $\theta_{13}$, and
one \CP{} violating phase $\delta_{\rm CP}$). This implies that
$\rho_i$ and $\tau_i$ would be equal if there was no mixing in the
leptonic sector. Furthermore, it is obvious that the matrices $\rho_i$
can be written as linear combinations of the matrices $\tau_i$ and
vice versa. Therefore, there is, in principle, no difference between
effects added in flavor or mass basis if one allows for the
most general form of the non-standard contribution.

We now define any non-standard effect as a ``pure'' flavor or mass
effect if the corresponding effective contribution to the Hamiltonian is
given by
\begin{equation}
H' = c \, \rho_{i} \quad {\rm or} \quad H' = c \, \tau_{i}, \quad
\mbox{($i$ fixed)}
\end{equation}
respectively, where $c \in \mathbb{R}$. This means that we
restrict the ``pure'' effects to be of very specific types, where the
actual forms are very simple in a given basis.\footnote{Because of our choice to use the Pauli matrices, a ``pure'' effect corresponds to the interaction of two flavor or mass eigenstates. This is also the reason for choosing to work with the Pauli matrices. In addition, it is also interesting to keep the real and complex parts of the off-diagonal entries separate (\ie, not working with the complex matrix elements directly, but rather a set of real parameters) in order to investigate the possibilities of probing \CP{} violation effects.} Given the
possible theoretical origin, this approach is quite plausible if one
assumes that the underlying theoretical model will produce one leading
flavor (or mass) changing or conserving effect.  Generally, the
parameter $c$ can depend on many different quantities, \eg, the matter
density or the neutrino energy. In particular, the dependence on the
neutrino energy (``spectral dependence'') may allow for the
unambiguous identification, or, in the case of mass-varying neutrinos,
the matter density dependence may indicate this type of
effect. However, any approach investigating such dependencies has to
use specific models, and the actual representation by Nature may
easily be overseen. Therefore, we do not require this information in
this study and rather investigate the generic impact of effects in the
flavor or mass basis. In addition, we note that the matter density or
energy dependence of the non-standard effects 
should be very weak for a given terrestrial
experiment with a specific matter density profile. Only for effects
motivated by MVN, \ie, mass effects, we will use the same energy
dependence as for the masses themselves for numerical simulations. In general, if a large span of energies is available, one should of course also try to distinguish different specific models through their different energy dependencies.

This choice of pure effects implies that only one of the generators of
the Lie algebra is present, since a general linear combination, such
as \equ{superposition}, can always be interpreted in both bases. Thus,
we define a flavor or mass conserving (violating) effect as any effect
where the effective contribution to the Hamiltonian is diagonal
(off-diagonal) in the corresponding basis.\footnote{Strictly speaking,
our definition distinguishes (in two-flavors) off-diagonal additions
proportional to $\lambda_1$ (real) or $\lambda_2$ (complex).} We note
that a pure flavor (mass) violating effect corresponds to some
interaction between two flavor (mass) eigenstates. For example, the
$su(2)$ generators $\rho_1$ and $\rho_2$ correspond to flavor
violating (or changing) effects, whereas $\rho_3$ corresponds to
flavor conserving effects. In summary, if we detect an arbitrary
non-standard effect, it is the simple form in flavor or mass basis
which makes it a flavor or mass effect by our definition. This
approach can be justified by the fact that the simplest models for
non-standard effects from the underlying theory correspond to specific
patterns for the effective addition to the Hamiltonian.
Therefore, our definition of a ``pure'' effect
is a conceptually new one and it refers to a class of effects, which can be
interpreted in different ways. However, since simplicity is a basic concept in
physics, this concept allows the choice of the most ``natural'' non-standard effects
for further testing.

The case of non-standard Hamiltonian effects on three-flavor neutrino
oscillations, \ie, the case of $n=3$, is quite similar to the one
described above for the two-flavor case. Instead of the Pauli
matrices, which are a basis of the $su(2)$ Lie algebra, we now have to
use the eight Gell-Mann matrices, which span the $su(3)$ Lie
algebra. Out of the Gell-Mann matrices, three are off-diagonal with
two real entries, three are off-diagonal with two imaginary entries, and two
are diagonal with real entries. Even though the principle of the
three-flavor case is the same as that of the two-flavor case, it
introduces many more parameters (more leptonic mixing angles, the
complex phase in the leptonic mixing matrix, the extra mass squared
difference, and the extra degrees of freedom for the non-standard
effects), and therefore, turns out to be much more cumbersome to
handle than the two-flavor case. In the following, we will
therefore start by treating the two-flavor case in some detail and
then continue by studying the similarities and differences when
approaching the full three-flavor case.

As far as the classification of current models in our notation
is concerned, NSI and FCNC will be flavor effects, whereas MVN
will produce mass effects. In general, NSI can be of two types: flavor
changing (FC) and non-universal (NU) \cite{Valle:2003uv}. The
off-diagonal elements of the effective NSI potential
$\epsilon_{\alpha\beta}$, where $\alpha \neq \beta$, correspond to FC,
whereas the differences in the diagonal elements
$\epsilon_{\alpha\alpha}$ correspond to NU. In addition, FCNC are
flavor violating effects and MVN can be mass conserving. In
principle, for our purposes, there is no difference between FC NSI and
FCNC.

\section{Non-standard Hamiltonian effects in the two-flavor limit}
\label{sec:nonstandard_2}

In this section, we study the general implications of
non-standard Hamiltonian effects in the two-flavor limit. 
We discuss the effective parameter mapping including
non-standard effects, and then, we apply it to a two-flavor
limit as an example.

\subsection{Parameter mapping in two flavors}
\label{sec:pm}

In \App~\ref{app:twoflavor}, we
describe the general formalism of the two-flavor scenario, which can
be used to obtain the results in this section.
First, we discuss effects given in flavor basis, which are effects
expanded in $\rho_i$ [\cf, \equ{sigma_tau}]. In this case, flavor conserving effects will be
contributions to the total Hamiltonian on the form $H' = F_3 \rho_3$,
where $F_3 \in \mathbb R$, whereas flavor violating effects will be
contributions on the form $H' = F_1 \rho_1 + F_2 \rho_2$, where $F_i
\in \mathbb R$. In flavor basis, the new effective parameters
are given by
\begin{align}
\Delta \tilde m^2 &=
\Delta m^2 \xi, \label{equ:dmflavor} \\
\sin^2(2\tilde\theta) &=
\frac{\left[
\frac{4E F_1}{\Delta m^2} + \sin(2\theta)
\right]^2
+ \left(\frac{4E F_2}{\Delta m^2}\right)^2}{\xi^2}, \label{equ:thetaflavor}
\end{align}
where 
\begin{equation}
\xi = \sqrt{
\left[
\frac{4E F_1}{\Delta m^2} + \sin(2\theta)
\right]^2
+ \left(\frac{4E F_2}{\Delta m^2}\right)^2
+
\left[
\frac{2VE}{\Delta m^2}
+ \frac{4E F_3}{\Delta m^2} - \cos(2\theta)
\right]^2
} \label{equ:xiflavor}
\end{equation} 
is the normalized length of the Hamiltonian vector (see
\App~\ref{app:twoflavor}), $\Delta \tilde m^2$ is the effective mass
squared difference in flavor basis, and $\tilde \theta$ is the
effective mixing angle in flavor basis.\footnote{Note that $F_2$ may
also change the effective Majorana phase.} In addition, the resonance
condition is found to be
\begin{equation}
\frac{2VE}{\Delta m^2}
+ \frac{4E F_3}{\Delta m^2} = \cos(2\theta),
\label{equ:rescondflavor}
\end{equation}
which is clearly nothing but a somewhat modified version of the
Mikheyev--Smirnov--Wolfen\-stein (MSW) resonance condition
\cite{Wolfenstein:1977ue,Mikheev:1986gs,Mikheev:1986wj}. {}From the
resonance condition in \equ{rescondflavor}, it is easy to observe that
the resonance is present for some energy $E$ if and only if
\begin{equation}
\sgn(\Delta m^2) \, \sgn(V) \, \sgn(1+2F_3/V) = \sgn[\cos(2\theta)],
\label{equ:ressign}
\end{equation}
where $\sgn(\Delta m^2)$ is dependent on the mass hierarchy, $\sgn(V)$
is dependent on if we are studying neutrinos or anti-neutrinos, and
$\sgn(1+2F_3/V)$ is dependent on the ratio between $F_3$ and the
matter potential $V$ [$\sgn(1+2F_3/V)$ being equal to $-1$ if and only
if $F_3$ has a magnitude larger than $|V/2|$ and is of opposite sign
to $V$]. Note that if there are flavor violating contributions added
to the Hamiltonian, then these do not change the resonance
condition. The sign of $\cos(2\theta)$ can be made positive by reordering the mass eigenstates in the case of two neutrino flavors. However, we keep this term as it is, since this is not possible in the case of three neutrino flavors. This resonance condition can be easily understood, since the effective
contribution to the Hamiltonian from any flavor violating effect will
be parallel to the $H_3 = 0$ plane, \ie, these contributions are
off-diagonal.

If we choose to describe the non-standard addition to the
Hamiltonian in the mass eigenstate basis, then we find that the mixing
parameters are given by
\begin{align}
\Delta \tilde m^2 &= \Delta m^2 \xi, \\
\sin^2(2\tilde\theta) &=
\frac{\left[
\frac{4E M_1}{\Delta m^2}\cos(2\theta) +
\left(1 - \frac{4E M_3}{\Delta m^2}\right) \sin(2\theta)
\right]^2
+ \left(\frac{4E M_2}{\Delta m^2}\right)^2
}{\xi^2}, \label{equ:thetamass}
\end{align}
where
\begin{equation}
\xi =
\sqrt{
\left[
\frac{2VE}{\Delta m^2}\sin(2\theta) + \frac{4E M_1}{\Delta m^2}
\right]^2
+
\left(
\frac{4E M_2}{\Delta m^2}
\right)^2
+
\left[
\frac{2VE}{\Delta m^2}\cos(2\theta) + \frac{4 E M_3}{\Delta m^2} - 1
\right]^2
}, \label{eq:ximass}
\end{equation}
and the resonance condition becomes
\begin{equation}
\frac{2VE}{\Delta m^2} + \frac{4E M_1}{\Delta m^2}\sin(2\theta) +
\frac{4E M_3}{\Delta m^2}\cos(2\theta) = \cos(2\theta).
\end{equation}
Note that both mass conserving effects and mass violating effects
enter into the resonance condition, whereas only the flavor conserving
effects entered in the corresponding expression in the flavor basis
[\cf, \eq~(\ref{equ:rescondflavor})]. This is due to the fact that the
changes of the Hamiltonian vector from such effects are not parallel
to the $H_3 = 0$ plane (in flavor basis, see
\App~\ref{app:twoflavor}), \ie, both of these effects affect the
diagonal terms of the total Hamiltonian. However, $M_2$ does not enter
into the resonance condition, since $\sigma_2 = \tau_2$, \ie, the
change of the Hamiltonian is off-diagonal also in the flavor basis.

\subsection{Interpretation of experiments in the two-flavor limit}
\label{sec:exp}

Since a general analytic discussion of three-flavor neutrino
oscillations including non-standard Hamiltonian effects would be very
complicated, we focus on two neutrino flavors in this section. 
This approach can be justified if one assumes that the other contributions are exactly
known or the two-flavor probabilities dominate. Of course, for
short-term applications, small non-standard effects might be confused
with other small effects such as $\stheta$
effects~\cite{Huber:2002bi}. Thus, a comprehensive quantitative
discussion would be very complicated at present.

In three-flavor neutrino oscillations, we can construct several
interesting two-flavor limits of the probabilities $P_{\alpha \beta}$
including non-standard effects related to two-flavor neutrino
oscillations (see, \eg, \Ref~\cite{Akhmedov:2004ny}):
\begin{eqnarray}
P_{ee} & \underset{ \sdm \rightarrow 0}{\longrightarrow } & 1 - \sin^2
(2 \tilde{\theta}_{13}) \sin^2 \left( \frac{\Delta \tilde{m}_{31}^2
L}{4 E} \right), \label{eq:Pee_limit}\\
P_{ee} & \underset{ \theta_{13} \rightarrow 0}{\longrightarrow } & 1 -
\sin^2 (2 \tilde{\theta}_{12}) \sin^2 \left( \frac{\Delta
\tilde{m}_{21}^2 L}{4 E} \right), \\
P_{\mu e} & \underset{ \sdm \rightarrow 0 }{\longrightarrow } &
\sthetat \, \sin^2 \left( \frac{\Delta \tilde{m}_{31}^2 L}{4 E}
\right) \, \sin^2(\theta_{23}), \label{equ:Pemu} \\
P_{\mu\mu} & \underset{ \sdm \rightarrow 0, \, \theta_{13} \rightarrow
0}{\longrightarrow } & 1 - \sin^2 (2 \tilde{\theta}_{23}) \sin^2
\left( \frac{\Delta \tilde{m}_{31}^2 L}{4 E} \right). \label{eq:Pmm_limit}
\end{eqnarray}
Note that all of these probabilities also contain the standard matter
effects except from $P_{\mu \mu}$.  In general, the $su(3)$ generators
(the Gell-Mann matrices) will give the degrees of freedom for
non-standard Hamiltonian effects with three flavors. However, when
studying the effective two-flavor neutrino oscillations, we only use
the Gell-Mann matrices which are the equivalents of the Pauli matrices
in the two-flavor sector that is studied. In addition, one can create
two-flavor limits for oscillations into sterile neutrinos, such as in
\Ref~\cite{Pas:2005rb}.
In the following, we will focus on small mixing and the case of \equ{Pemu}
for illustration. We discuss the large mixing case in \App~\ref{app:visualization}.
In addition, see \App~\ref{app:Pemu} for subtleties with the definitions of the effective two-flavor scenarios.

\begin{figure}
\includegraphics[width=\textwidth]{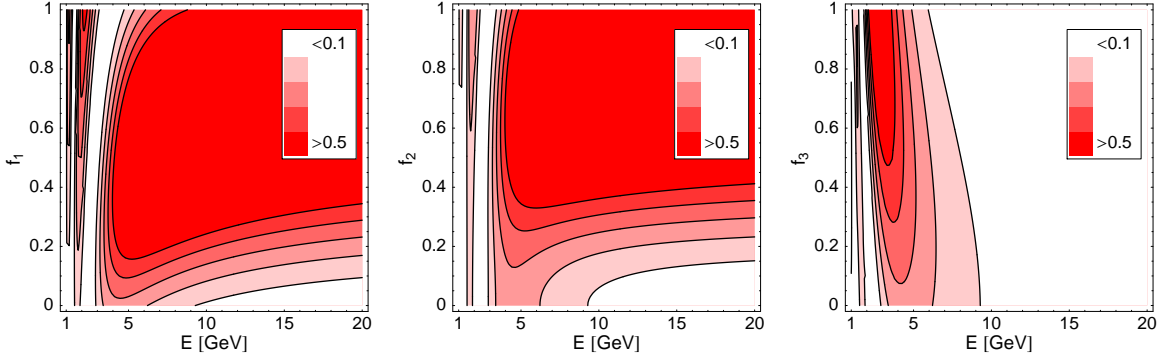}
\caption{\label{fig:feffects} The two-flavor appearance probability
    $P_{\alpha \beta}$ as a function of energy and the flavor
    conserving/violating fraction $f_i \equiv F_i/V$
    (normalized relative to matter effects). For the values of the
    neutrino parameters, we have used $\theta=0.16 \simeq 9.2^\circ$,
    $\Delta m^2 = 0.0025 \, \mathrm{eV}^2$, $L=3000 \, \mathrm{km}$,
    $\rho = 3.5 \, \mathrm{g}/{\mathrm{cm}^3}$, neutrinos only,
    $\Delta m^2>0$, and $f_i>0$.}
\end{figure}

For small mixing, such as for \equ{Pemu}, we show in \figu{feffects}
the neutrino oscillation appearance probability $P_{\alpha\beta}$ for
two flavors with small mixing, where the effects of the $F_i$'s are
parameterized relative to the matter effects (\ie, ``1'' on the
vertical axis corresponds to an effect with $F_i = V$ and ``0'' to no
non-standard effects). In this figure, many of the following analytic
observations are visualized. The resonance condition in
\equ{rescondflavor} can always be fulfilled for the matter resonance
($F_3=0$) by an appropriate choice of energy, baseline, neutrinos or
antineutrinos, and oscillation channel. Obviously, we can read off
from \equ{thetaflavor} that at the resonance $\sin^2(2 \tilde{\theta})
\rightarrow 1$, where the matter resonance condition can be influenced
by $F_3$ according to \equ{rescondflavor}. Therefore, the magnitude of
$\sin^2(2 \tilde{\theta})$ at the resonance (but not necessarily
$P_{\alpha\beta}$) is independent of $F_1$, $F_2$, and $F_3$ by definition. However,
$F_3$ can shift the position of the resonance (such as in energy
space). If we choose an energy far above the resonance energy and
$F_i/V \ll 1$ ($i = 1,2,3$), then we have
\begin{equation}
\sin^2 (2 \tilde{\theta}) \rightarrow
\frac{\left[
\frac{4E F_1}{\Delta m^2} + \sin(2\theta)
\right]^2
+ \left(\frac{4E F_2}{\Delta m^2}\right)^2}{
\left[
\frac{2VE}{\Delta m^2}
+ \frac{4E F_3}{\Delta m^2} - 1
\right]^2
}.
\label{equ:offres}
\end{equation}

This means that $F_1$ and $F_2$ can, for large enough energies, enhance
a flavor transition, \ie, they increase the oscillation amplitude. 
In principle, one could
distinguish $F_1$ from $F_2$ by a measurement at two different
energies, because the mixed term from the square in the numerator of
\equ{offres} has a linear (instead of quadratic)
energy dependence. In practice, such a discrimination should be very
hard. In addition, the quantity $F_3$ can play the same role as the
matter potential $V$, \ie, it can change the flavor transition for
large energies. It is also obvious from
\eqs~(\ref{equ:thetaflavor}) and (\ref{equ:xiflavor})
 that $F_3$ can affect the matter resonance energy and that it 
is directly correlated with the matter
potential $V$, \ie, one cannot establish effects more precisely than
the matter density uncertainty.

In \Sec~\ref{sec:pm}, we have also discussed mass effects, such as
coming from MVN. Since a pure $M_1$ or $M_3$ effect translates into a
combination of $F_1$ and $F_3$ [\cf,
\equ{flavor_mass_transformation}], we expect to find a mixture of
$F_1$ and $F_3$ effects, \ie, both $F_1$ and $F_3$ effects have to be
present. Thus, if we assume that there is only one dominating ``pure''
non-standard contribution ($F_1$, $F_2$, $F_3$, $M_1$, $M_2$, or
$M_3$), then this simultaneous presence points toward a mass
effect. Clearly, an $M_2$ effect, on the other hand, cannot be
distinguished from an $F_2$ effect [\cf, \equ{flavor_mass_transformation}].
A different property of $M_3$, which is
not so obvious from \Sec~\ref{sec:pm}, but very obvious
already from \eqs~(\ref{equ:hamvac}), (\ref{equ:ns}), and
(\ref{equ:sigma_tau}): Since $M_3$ is diagonal in mass basis, it
corresponds to an energy dependent shift of the vacuum mass squared
difference. As a consequence, in vacuum, the effective mixing angle is
not modified by $M_3$ [\cf, \equ{thetamass}]. Thus, the oscillation
amplitudes are not modified by $M_3$, but the oscillation pattern
shifts (contrary to $F_3$ effects, where also the amplitude
changes). In this case, the resonance condition becomes meaningless
and the amplitude becomes $\sin^2 (2 \tilde{\theta} ) = \sin^2 ( 2
\theta)$. Note that a direct test using one experiment only makes
it hard to identify mass effects uniquely if they are introduced with
the same energy dependence as the vacuum masses (because they can be
rotated away by a different set of neutrino oscillation parameters). Thus, other 
methods might be preferable, such as modified MSW transitions in the
Sun~\cite{Barger:2005mn,Cirelli:2005sg} or reactor experiments
comparing air and matter oscillations~\cite{Schwetz:2005fy}). 

Another class of effects has been discussed by Blennow {\it et al.}~in
\Ref~\cite{Blennow:2005yk}. In this study, so-called ``damping
effects'' could describe modifications on probability level instead of
Hamiltonian level (such as neutrino decay, absorption, wave packet
decoherence, oscillations into sterile neutrinos, quantum decoherence,
averaging, \etc). It is obvious from \eq~(3) in
\Ref~\cite{Blennow:2005yk} that these damping effects do not alter the
oscillation frequency, while we can read off from
\eqs~(\ref{equ:dmflavor}) and (\ref{equ:xiflavor}) that it is a
general feature of non-standard Hamiltonian effects that the
oscillation frequency is changed. However, for damping effects, the
oscillation amplitude can be damped either by a damping of the overall
probability (``decay-like damping'') or by the oscillating terms only
(``decoherence-like damping''). In the first case, the total
probability of finding a neutrino in any neutrino state is damped for
all energies, whereas in the second case, it is constantly equal to
one while the individual neutrino oscillation probabilities are damped
in the oscillation maxima and enhanced in the oscillation minima.
Since all (small) effects one could imagine in quantum field theory,
involving the modification of fundamental interactions or
propagations, can be described by either coherent or incoherent
addition of amplitudes, one can expect that the two classes of
Hamiltonian and probability (damping) effects can cover all possible
effects. However, in practice, potential energy, environment, and
explicit time dependencies (such as from a matter potential) can make
life more complicated.

\section{Three-flavor effects}
\label{sec:threeflavtheory}

As was stated in the \Sec~\ref{sec:nonstandard_n}, the general
three-flavor case is quite complicated. However, if we assume that the
non-standard effects are small, then we can use perturbation theory to
derive expressions for the change in the neutrino oscillation
parameters. For example, the elements of the effective mixing matrix
are given by
\begin{equation}
	\tilde U_{\alpha i} = \braket{\nu_\alpha}{\tilde \nu_i},
\end{equation}
where $\ket{\tilde \nu_i}$ is the eigenstate of the full
Hamiltonian. To first order in perturbation theory, we have
\begin{equation}
	\ket{\tilde \nu_i} = \ket{\nu_i} + \sum_{j\neq i} 
	\frac{\obraket{\nu_j}{H'}{\nu_i}}{E_i - E_j} \ket{\nu_j}
	\simeq
	\ket{\nu_i} + 2E \sum_{j\neq i} \frac{H'_{ji}}{\Delta m_{ji}^2} \ket{\nu_j}, 
\end{equation}
and thus, we find
\begin{equation}
	\tilde U_{\alpha i} \simeq U_{\alpha i} + 
	2E \sum_{j\neq i} \frac{H'_{ji}}{\Delta m_{ji}^2} U_{\alpha j}
\end{equation}
or, in terms of the non-standard addition given in flavor basis,
\begin{equation}
	\tilde U_{\alpha i} \simeq U_{\alpha i} + 
	2E \sum_{j\neq i}\sum_{\beta,\gamma} \frac{U_{\beta j}U^*_{\gamma i} H'_{\beta\gamma}}
	{\Delta m_{ji}^2} U_{\alpha j}.
\end{equation}
We note that this approach is valid only if $|2EH'_{ij}/\Delta m_{ji}^2|
\ll 1$. If this is not valid, then we have to use degenerate
perturbation theory in order to obtain valid results.

It was discussed in \Refs~\cite{Huber:2001de,Huber:2002bi}, that if $\theta_{13}$ is small enough, then possible NSI in the creation, propagation, and detection processes may mimic the effects of a larger $\theta_{13}$ (this can also be the case for other effects
which are not usually treated along with neutrino oscillations, such
as damping effects \cite{Blennow:2005yk}). Here, we consider only the propagation effects separately and consider how this alone could affect the determination of $\theta_{13}$. The reason for doing so is that, while NSI can also affect the creation and detection processes, other non-standard effects, \eg, MVN, may not. With the perturbation theory approach described above, this becomes very transparent, and is probably one of the most interesting applications of non-standard effects. In any
experimental setup, the value of the mixing angle $\theta_{13}$ is
determined by the modulus of the element $U_{e3}$ of the neutrino
mixing matrix $U$. If we include non-standard effects, then the effective
counterpart of this element is given by
\begin{eqnarray}
	\tilde U_{e3} &\simeq& U_{e3} +
	\frac{2E}{\Delta m_{31}^2}(1+\alpha s_{12}^2)(s_{23}H'_{e\mu}+c_{23}H'_{e\tau}) +
	\nonumber \\
	&& \alpha \frac{2E}{\Delta m_{31}^2}
	s_{12}c_{12} \left[
	c_{23}^2 H'_{\mu\tau} - s_{23}^2H'_{\tau\mu} + 
	\frac 12 \sin(2\theta_{23})(H'_{\mu\mu} - H'_{\tau\tau})
	\right],
\end{eqnarray}
where we have made a series expansion to first order in $\alpha =
\Delta m_{21}^2/\Delta m_{31}^2 \simeq 0.03$ and disregarded terms of
second order in both $H'$ and $\theta_{13}$.

If $U_{e3}$ is smaller than or of equal size to the other terms in
this expression, then the $\theta_{13}$ determined by an experiment
will not be the actual $\theta_{13}$ unless the non-standard effects
are taken into account. It is worth to notice that if $\theta_{23} =
45^\circ$, then $c_{23} = s_{23}$ and only the imaginary part of
$H'_{\mu\tau} = (H'_{\tau\mu})^*$ will enter into the expression for
$\tilde U_{e3}$, indicating that if the leading term is the one
containing $H'_{\mu\tau}$, then the effective \CP{} violating phase will
be $\pm 90^\circ$. Another interesting observation is that even if
there are no non-standard effects, there is a term proportional to
$\Delta V \equiv H'_{\mu\mu} - H'_{\tau\tau}$ in this
expression. Because of the different matter potentials for $\nu_\mu$
and $\nu_\tau$ due to loop-level effects, this quantity will be of the
order $\Delta V \simeq 10^{-5} V$.

\begin{figure}[t]
\begin{center}
\includegraphics[width=12cm]{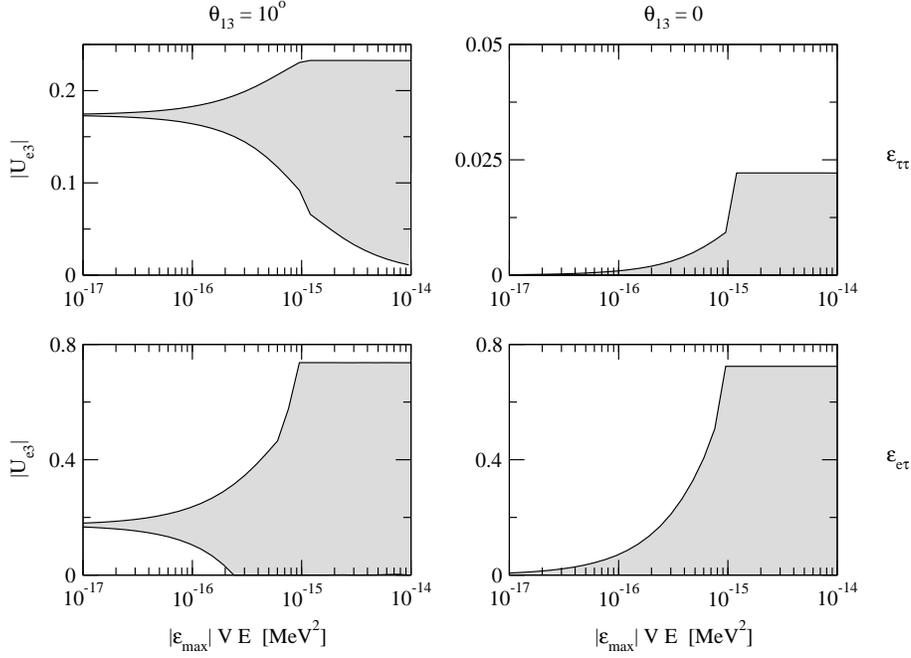}
\caption{The range of possible $|\tilde U_{e3}|$ as a function of
$\eps_{\rm max}VE$. The plots are arranged so that the left panels
correspond to $\theta_{13} = 10^\circ$ and the right panels to
$\theta_{13} = 0$, while the lower panels correspond to a non-standard
effect with $\eps_{e\tau} \neq 0$ and the upper panels to a
non-standard effect with $\eps_{\tau\tau} \neq 0$. The qualitative
behavior for other values of $\theta_{13}$ is similar to the behavior
for $\theta = 10^\circ$. (Note the different scales on the vertical
axes.)}
\label{fig:threeeff}
\end{center}
\end{figure}

In \fig~\ref{fig:threeeff}, we plot the possible range of $|\tilde
U_{e3}|$ as a function of $ \eps_{\rm max}VE$, where $H'_{\alpha\beta}
= \eps_{\alpha\beta} V$, $V$ is the matter potential, and
$|\eps_{\alpha\beta}| < \eps_{\rm max}$. For comparison, a neutrino
factory with a neutrino energy of $E = 50$ GeV and a matter density of
3 g/cm$^3$ will have $VE \simeq 6\cdot 10^{-15}$ MeV$^2$ and the
position at which we need to consider the possible range of $|\tilde
U_{e3}|$ then depends on the bounds on the non-standard
parameters $\eps_{\alpha\beta}$. In general, the bounds for
$\eps_{\alpha\beta}$ depend on the type of non-standard effect and the types of interactions that are
considered. In the case of NSI, it is common to write the non-standard interaction
parameters as
\begin{equation}
\eps_{\alpha\beta} = \sum_f \eps^f_{\alpha\beta} \frac{N_f}{N_e},
\end{equation}
where we sum over different types of fermions, $\eps^f_{\alpha\beta}$
depends on the non-standard interaction with the fermion $f$, and
$N_f$ is the number density of the fermion $f$. In addition,
$\eps^f_{\alpha\beta}$ is often split into $\eps^f_{\alpha\beta} =
\eps^{fL}_{\alpha\beta} + \eps^{fR}_{\alpha\beta}$, where $L$ and $R$
denotes the projector used in the fermion factor of the effective
non-standard Lagrangian density, \ie,
\begin{equation}
	\mathcal{L}_{\rm eff} = - 2\sqrt 2 G_F \sum_{f} \sum_{P = L,R} \eps^{fP}_{\alpha\beta} (\bar\nu_\alpha\gamma_\rho L \nu_\beta)(\bar f \gamma^\rho P f). 
\end{equation}
Recent bounds for $\eps^{fP}_{\alpha\beta}$ can be found in
\Ref~\cite{Barranco:2005ps} for electron neutrino interactions with
electrons (\ie, $\eps^{eP}_{e\beta}$) and \Ref~\cite{Davidson:2003ha} for
interactions with first generation Standard Model fermions. As an
example, the bounds from \Ref~\cite{Barranco:2005ps} for the $\eps_{e\tau}$
(which is considered in \fig~\ref{fig:threeeff}) are
\begin{eqnarray}
	-0.90 < \eps^{eL}_{e\tau} < 0.88 & \quad {\rm and} \quad &
	-0.45 < \eps^{eR}_{e\tau} < 0.44,
\label{equ:bounds}
\end{eqnarray}
respectively. This means that the bounds, especially in this sector, are
weak, which we will use in the next section.
{}From \fig~\ref{fig:threeeff}, we can deduce that the off-diagonal
$\eps_{e\tau}$ terms have a larger potential of altering the value of
$|\tilde U_{e3}|$ than the diagonal $\eps_{\tau\tau}$ terms, the
maximal value can even exceed $1/\sqrt 2$, corresponding to $\tilde
\theta_{13} = 45^\circ$. In addition, it is possible to suppress the
effective $\theta_{13}$ to zero if introducing non-standard
effects. It follows that a relatively large $\theta_{13}$ signal,
bounded only by the size of the non-standard effects, can be induced
or that a large $\theta_{13}$ signal can be suppressed by non-standard
effects.  Note that the effects quickly disappear at low energies, \eg, in reactor experiments. In
order to tell a genuine $\theta_{13}$ signal apart from a signal
induced by non-standard interactions, it is necessary to study the
actual distortion of the energy spectrum induced by the neutrino
oscillations.

\section{A numerical example: Neutrino factory for large $\boldsymbol{\stheta}$}
\label{sec:numex}

This section is not supposed to be a complete study of non-standard
Hamiltonian effects, but to demonstrate some of the qualitatively
discussed properties from the last sections in a complete numerical
simulation of a possible future experiment using the exact three-flavor
probabilities. Therefore, we have to make
a number of assumptions. We use a modified version of the GLoBES
software~\cite{Huber:2004ka} to include non-standard effects. As a
future high-precision instrument, we choose the neutrino factory
experiment setup from \Refs~\cite{Huber:2002mx,Huber:2005jk} with $L=3
\, 000 \, \mathrm{km}$, a $50 \, \mathrm{kt}$ magnetized iron
calorimeter detector, $1.06 \cdot 10^{21}$ useful muon decays per
year, and four years of running time in each
polarity.\footnote{Compared to \Ref~\cite{Huber:2002mx}, we use a
2.5~\% systematic normalization error for all channels as in
\Ref~\cite{Huber:2005jk}.}  This experiment uses muon neutrino disappearance
and electron to muon neutrino appearance as oscillation channels for both
neutrinos and antineutrinos (in the muon and anti-muon operation modes combined).
For the neutrino oscillation parameters,
we use $\sin^2 2 \theta_{12}=0.83$, $\sin^2 2 \theta_{23} = 1$, $\sdm
= 8.2 \cdot 10^{-5} \, \mathrm{eV}^2$, and $\ldm = 2.5 \cdot 10^{-3}
\,
\mathrm{eV}^2$~\cite{Fogli:2003th,Bahcall:2004ut,Bandyopadhyay:2004da,Maltoni:2004ei},
as well as we assume a 5~\% external measurement for $\sdm$ and
$\theta_{12}$ \cite{Bahcall:2004ut} and include matter density
uncertainties of the order of
5~\%~\cite{Geller:2001ix,Ohlsson:2003ip}. In order to test precision
measurements of the non-standard effects, we use 
$\stheta=0.1$ close to the CHOOZ upper bound\footnote{In general, a large $\stheta$ will imply a large signal in the appearance channel. However, non-zero effective $\stheta$ could arise even if $\stheta = 0$, \cf, \fig~\ref{fig:threeeff}. For effects which are diagonal in flavor basis, a large $\stheta$ would be preferred in order to make an observation of the non-standard effect. We have used large $\stheta$ as an example, since one may argue that the finding of new effects at present experiments (such as MINOS) may lead to a good reason for constructing a neutrino factory. One should also observe that, in principle, it would be possible to find non-standard effects at, \eg, MINOS \cite{Kitazawa:2006iq,Friedland:2006pi}. However, the precision of a neutrino factory would be more sensitive to small effects, and thus, more useful for distinguishing between effects.}~\cite{Apollonio:1999ae}, as well as
we assume a normal mass hierarchy and $\deltacp=0$. For simplicity, we
do not take the $\mathrm{sgn}(\ldm)$-degeneracy~\cite{Minakata:2001qm}
into account, but we include the intrinsic
$(\theta_{13},\deltacp)$-degeneracy~\cite{Burguet-Castell:2001ez},
whereas the octant degeneracy does not appear for maximal
mixing~\cite{Fogli:1996pv}. Note that we do {\em not} include external
bounds on the non-standard physics and $\stheta$, which, for instance,
mean that we allow ``fake'' solutions of $\stheta$ above the CHOOZ
bound. This assumption is plausible, since, depending on the effect,
the CHOOZ bound may have been affected by the non-standard effect as
well.

\subsection{Test model}

Since we choose $\stheta$ to be large, let us first of all focus on the appearance
channel of $\nu_e$ oscillating into $\nu_\mu$ (or $\bar\nu_e$
oscillating into $\bar\nu_\mu$). Expanding in small $\stheta$ and
$\alpha \equiv \sdm/\ldm$, we have for $\alpha \rightarrow 0$ (which
should be a good approximation for $\stheta \gg \alpha^2 \simeq
0.001$)~\cite{Cervera:2000kp,Freund:2000ti,Freund:2001pn}
\begin{equation}
P_{e\mu} \sim  \sin^2 2 \theta_{13}  \, \sin^2 \theta_{23} \frac{\sin^2[(1-\hat{A}){\Delta}]}{(1-\hat{A})^2},
\label{equ:Papp}
\end{equation}
where $\Delta \equiv \Delta m_{31}^2 L/(4 E)$ and $\hat{A} \equiv \pm
2 \sqrt{2} G_F n_e E/\Delta m_{31}^2$. Similarly, $P_{e\tau}$ is described by this equation
with $\sin^2 \theta_{23}$ replaced by $\cos^2 \theta_{23}$. This means that we may be
effectively dealing with the two-flavor limits described in
\Sec~\ref{sec:nonstandard_2}, depending on the degree the non-standard effects are different for the  
$\mu$ and $\tau$ flavors (\cf, \App~\ref{app:Pemu}).

Using the parameterization in \eqs~(\ref{equ:superposition}) and (\ref{equ:pauli})
applied to the 1-3-sector, we therefore adopt the following
Hamiltonian:
\begin{eqnarray}
H_{\rm eff} & = & \frac{1}{2 E} U \left( 
\begin{array}{ccc}
\tilde{M}_3 & 0 & \tilde{M}_1 - {\rm i} \tilde{M}_2 \\
0 & \Delta m_{21}^2 & 0 \\
\tilde{M}_1+{\rm i} \tilde{M}_2  & 0 & \Delta m_{31}^2 - \tilde{M}_3 
\end{array}
\right) U^\dagger  \nonumber \\
& + & \left( 
\begin{array}{ccc}
V+F_3 & 0 & F_1 - {\rm i} F_2 \\
0 & 0 & 0 \\
F_1+{\rm i} F_2  & 0 & - F_3  \\
\end{array}
\right) \,  .
\label{equ:numham}
\end{eqnarray}
In this model, $M_1$ and $M_2$ correspond to the \CP{} conserving and \CP{} violating parts of a
mass-changing effect, whereas $M_3$ is a mass-conserving effect. In addition, $F_1$ and $F_2$ are the \CP{} conserving and \CP{} violating parts of a
flavor-changing effect, whereas $F_3$ is a flavor-conserving effect. As motivated before, it is plausible to assume that one of these non-standard effects may be dominating the other ones, because many models predict such a dominating component and the experimental constraints on some quantities are rather strong.
In addition, \equ{numham} implies that the effects are mainly present in the 1-3-sector, which can be motivated by rather weak experimental bounds on the $\nu_\tau$-sector. For example, the bounds on the matrix element $H'_{e\tau}$ are rather weak in the case of NSI, making it viable that this term is dominating the NSI Hamiltonian. In this case, we obtain
\begin{equation}
H' = V \mtrx{ccc}{0 & 0 & \eps_{e\tau} \\ 0 & 0 & 0 \\ \eps_{e\tau}^* & 0 & 0} \quad
\Leftrightarrow \quad F_1 = V \real \eps_{e\tau}, \ F_2 = - V\imag \eps_{e\tau}.
\end{equation}
Thus, we have a flavor violating effect with $F_1$ representing the \CP{} conserving part of the NSI and $F_2$ representing the \CP{} violating part of the NSI.
 The form of the mass effects has been chosen to match the expected energy dependence of MVN in order to discuss effects with realistic spectral (energy) dependencies.

Note that the parameterization in \equ{numham} does not exactly correspond to the
two-flavor limit even for $\alpha \rightarrow 0$, since there are some
non-trivial mixing effects in the 2-3-sector as described in
\App~\ref{app:Pemu}. This parameterization is also obviously not the whole story in the three-flavor scenario.  For instance, we assume the same sign for effects on neutrinos and antineutrinos, which may, depending on the model, not apply in general. 
 However, we will demonstrate some of
the characteristics from \Sec~\ref{sec:exp} with this approach. In
addition, note that we have now adopted a specific energy dependence of
the flavor and mass effects, where the definition of the energy dependence in the $\tilde{M}$'s is slightly different from the one in the $M$'s in \Sec~\ref{sec:nonstandard_n}, \ie, $M \equiv \tilde{M}/(2E)$. In this case, the mass effects could be coming from MVN changing the mass eigenstates, whereas the
flavor effects correspond to some NSI
approximately constant in the considered energy range. We will quantify the
size of the $F_i$ and $\tilde{M}_i$ in terms of the normalized quantities $f_i \equiv F_i/V$ (for $\rho = 3.5 \,
\mathrm{g/cm^3}$) and $\mu_i \equiv \tilde{M}_i/\ldm$ (for $\ldm = 2.5 \cdot
10^{-3} \, \mathrm{eV}^2$). This quantification makes sense, since it
is obvious from \equ{numham} that the effect of these quantities will
have to be compared with the order of $V$ and $\ldm$, respectively.
Note that $f_1 - {\rm i} f_2 = \epsilon_{e \tau}^e$ from
\Sec~\ref{sec:threeflavtheory}, which means that it will be interesting
to compare the precisions of $f_1$ and $f_2$ to the current bounds
for $\epsilon_{e \tau}$. Furthermore, note that the mass effects
can be simply rotated away by a different choice of the mixing matrix
and the mass squared differences because of the same energy dependence in this example.
However, since we assume the solar parameters to be measured externally, we will observe that constraints to the $\tilde{M}_i$ can be derived. Such an external measurement with a similar environment
dependence to the neutrino factory comes from KamLAND, which turns out to be very
consistent with the ones from solar neutrino experiments. Since most non-standard effects in
oscillations are dependent on the matter density (such as MVNs with acceleron couplings to matter fields,
or non-standard flavor-changing matter effects generated by higher-dimensional operators), it is plausible  to assume that strong constraints hold for the solar sector because of the very different environments/densities within the Sun and the Earth.

\subsection{Identifying specific pure effects}

\begin{figure}[t]
\begin{center}
\includegraphics[width=8cm]{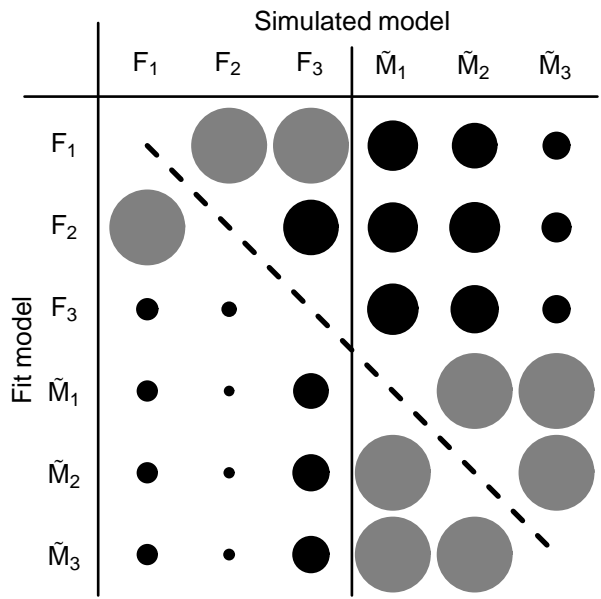}
\end{center}
\caption{\label{fig:corrmatrix} Correlation between simulated models
(columns) and fit models (rows). The areas of the disks represent the
discovery potentials of the simulated ``pure'' effects (parameterized
in terms of $f_i$ or $\mu_i$) given that a different pure
effect (fit model) is allowed (minimum value of a deviation
from zero necessary in either direction for a 3$\sigma$
discovery). Therefore, the larger the disk, the more difficult it will
be to distinguish a pure effect from another one.  Note that we use
cutoffs of $|f_i| \lesssim 0.3$ and $|\mu_i| \lesssim 0.5$ (largest
gray disks), since some models cannot even be distinguished for much
larger values.  The areas of the rest of the disks are normalized with
respect to these cutoffs for simulated flavor and mass effects.  }
\end{figure}

If we discover a non-standard effect, it will be an interesting
question how easily it can be identified. Assuming one dominating
effect of the mass or flavor type, which we have introduced as ``pure
effect'', we want to know how well it can be distinguished from
other such effects of different qualitative nature.
Therefore, in \figu{corrmatrix}, we show the correlation
between simulated and fit pure effects. For this figure, we simulate a pure
effect (column) and fit it with a different one (row), \ie, we
marginalize over the respective $f_i$ or $\mu_i$.  The areas of the
disks are proportional to the minimum simulated value necessary to
establish a $3 \sigma$ effect, where we have chosen a cutoff of $|f_i|
\lesssim 0.3$ and $|\mu_i| \lesssim 0.5$ (corresponding to the largest
gray disks).\footnote{Note that, for instance, the gray disks for $f_1$ and $f_2$ 
correspond to the order of magnitude of the upper bounds in \equ{bounds}, which means 
that testing considerably larger effects does not make sense.}
 This means that the size of the disks measures the
correlation between two pure effects and the ability to discriminate
those. 

One can easily make a number of qualitative observations from
\Sec~\ref{sec:exp} quantitative. First, it is hard to discriminate
between $F_1$ and $F_2$ (\CP{} conserving and \CP{} violating flavor-changing
effects), since these effects are qualitatively similar
and highly correlated with $\theta_{13}$ (as we have tested). However,
if Nature implemented a flavor-changing $F_1$ or $F_2$ effect, then
one could easily establish it against $F_3$ and the pure mass
effects. In general, note that a discrimination between flavor and
mass effects is rather easy because of their different spectral
dependence in this example (such as between $F_2$ and $\tilde{M}_2$).
The difference to $F_3$ can be explained
by the different flavor-conserving nature of $F_3$. The results look
somewhat different for the $F_3$ column: Because of the correlation
with $\rho$ and all of the neutrino oscillation parameters (see
below), it will be hard to establish this effect. For the simulated 
mass effects, the scale is different, \ie, one cannot directly compare
the $\tilde{M}$-columns with the $F$-columns.  Again, the mass effects can be
distinguished from the pure flavor effects to some extent. However, it is
quite impossible to establish a mass effect against another one,
since they can be easily simulated by a different set of mass squared
differences and mixing parameters with the same energy dependence.
The only reason why the pure mass effects can be established in
this example at all is that we have imposed external constraints on the solar
parameters as motivated above.

\subsection{Discovery of non-standard physics and potential for improvements}

\begin{table}[t]
\begin{center}
\begin{tabular}{lrrrr}
\hline
Quantity & Lower limit ($1\sigma$) & Upper limit ($1\sigma$) & Lower limit ($3\sigma$) & Upper limit ($3\sigma$) \\
\hline
$f_1$ & $-$0.008 & 0.008 & $-$0.025 & 0.026 \\
$f_2$ & $-$0.003 & 0.003 & $-$0.008 & 0.008 \\
$f_3$ & $-$0.016 & 0.016 & $-$0.049 & 0.082 \\
$\mu_1$ & $-$0.176 & 0.118 & $-$0.218 & 0.211 \\
$\mu_2$ & $-$0.105 & 0.126 & $-$0.181 & 0.212 \\
$\mu_3$ & $-$0.015 & 0.015 & $-$0.044 & 0.090 \\
\hline
\end{tabular}
\end{center}
\caption{\label{tab:limits} Discovery limits for the parameters in \equ{numham} as parameterized $f_i = F_i/V$  and $\mu_i = \tilde{M}_i/\ldm$ from the neutrino factory simulation (including correlations).}
\end{table}

A very important issue of any pure non-standard effect is its evidence compared to 
the standard three-flavor oscillation scenario. Therefore, in \Tab~\ref{tab:limits}, we show the discovery reaches for the parameters from 
\equ{numham} against the standard three-flavor neutrino oscillation scenario. This means that
the shown pure effects are simulated and the standard three-flavor neutrino oscillation parameters are
marginalized over.  Comparing the precisions of $f_1$ and $f_2$ with the numbers in \equ{bounds} is impressive.
However, these discovery reaches depend on $\stheta$ (and $\deltacp$) and we have assumed a 
very large $\stheta=0.1$ (and $\deltacp=0$). Note that the reach in $f_2$ is actually better than
the one in $f_1$, which is different from what is found in the two-flavor limit in \Sec~\ref{sec:exp}. The reasons are the mixing effects in the 2-3-sector and that
$F_2$ is a non-trivial source of \CP{} violation in the three-flavor case.

\begin{figure}[t]
\begin{center}
\includegraphics[width=0.8\textwidth]{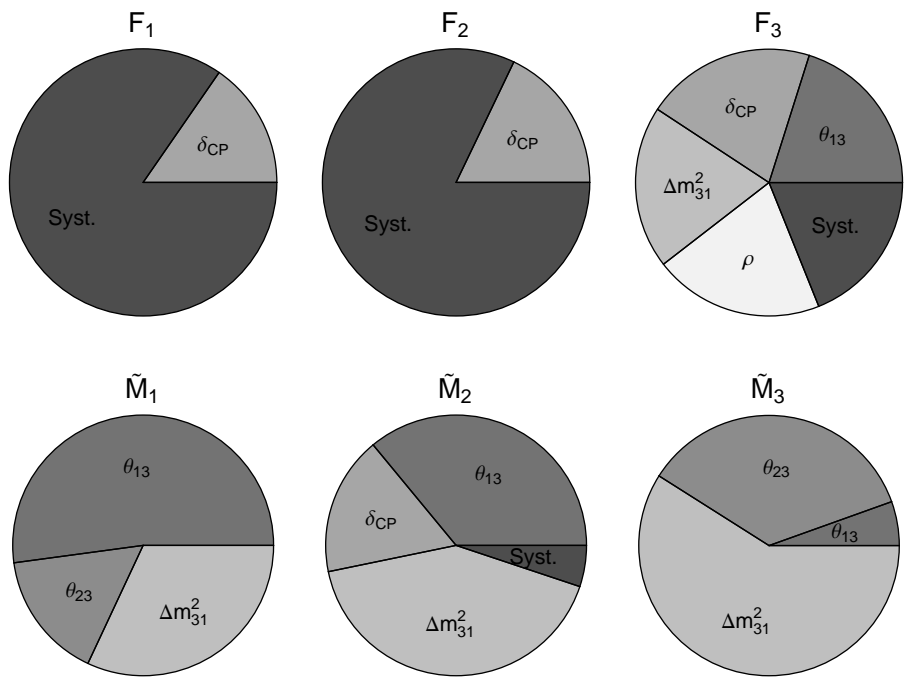}
\end{center}
\caption{\label{fig:impacts} Main impact factors (impact greater than
5~\%) for the test of specific simulated models (captions) against
standard three-flavor neutrino oscillations ($3 \sigma$
measurement). The neutrino oscillation parameters refer to
correlations with the respective parameter, ``Syst.'' refers to
systematics, and ``$\rho$'' refers to the matter density uncertainty.
The impact factors are defined as in \Ref~\cite{Huber:2002mx} as
relative improvement when the respective quantity is fixed
(correlations) or systematics is switched off.  }
\end{figure}

Except from these sensitivities, which somewhat depend on the specific model,
the behavior for neutrinos and antineutrinos, and so on, it may
be of some interest to obtain hints how these reaches can be improved. In order
to study this aspect, we show the so-called ``impact factors'' for the 
test of specific simulated models against standard three-flavor neutrino
oscillations in \figu{impacts}.  These impact factors test the relative impact of the
measurement errors on the neutrino oscillation parameters and systematics. In order to
compute them, the non-standard discovery limits are evaluated with all neutrino oscillation parameters
marginalized over, matter density uncertainties included, and systematics 
switched on (standard). 
In addition, in order to test a specific impact factor, one neutrino oscillation parameter
is fixed at one time (or systematics is switched off), 
and the corresponding discovery reach for the non-standard
effect is compared to the discovery reach including all uncertainties and systematics.
The difference between these to discovery reaches describes the impact of a particular
measurement error (or systematics), and the relative impact in \figu{impacts} quantifies
what one needs to optimize for in order to improve the discovery reach.
For example, for $\tilde{M}_3$ (lower right pie), the error on $\ldm$ is the
main impact factor in our model, which needs to be improved to increase the
 $\tilde{M}_3$ discovery reach.

Again, a number of aspects from \Sec~\ref{sec:exp} can
be verified. For $F_1$ and $F_2$ effects, systematics is the main
impact factor, since these flavor effects determine the overall height
of the appearance signal and are not introduced with a specific
spectral dependence (remember that we use a conservative overall
normalization error of 2.5~\%). For $F_3$ effects, we have earlier 
determined the matter density uncertainty
as an important constraint. However, improving the knowledge on
$\ldm$, $\theta_{13}$, or $\deltacp$ does have a similar effect, since
the extraction of the individual parameters becomes easier. For the
mass effects, we encounter a completely different behavior. Remember
that we have defined the mass effects with the same energy dependence
as the mass squared differences, which means that particularly $\tilde{M}_3$
is easily mixed up with $\ldm$. On the other hand, $\tilde{M}_1$ and $\tilde{M}_2$ are
related to a flavor change in the appearance channel via the mixing
matrix, \ie, the leptonic mixing angle $\theta_{13}$. Therefore, it is not surprising that such
a flavor change can be interpreted as either a mixing or a mass-changing
effect. Compared to \Sec~\ref{sec:exp}, there are also a number of
differences coming from the three-flavor treatment (solar and \CP{} effects) and the mixing in the 2-3-sector. These
effects introduce additional correlations with $\theta_{23}$ and
$\deltacp$.  However, they are also the reason why, for example, $\tilde{M}_3$ can be constrained
at all from this experiment alone [in the pure two-flavor case or without external constraints
on the solar parameters, it would be impossible to
distinguish between a non-vanishing $\tilde{M}_3$ and a different $\ldm$ if the mass effects had the energy dependence assumed in \equ{numham}].

\subsection{Comparison to damping effects}
\label{sec:comptodamp}

In the context of the non-standard effect identification, a more general question is the
ability to distinguish Hamiltonian effects and effects on probability level. The probability level effects lead to damping of the neutrino oscillation probabilities (``damping effects'') and
were studied in detail in \Ref~\cite{Blennow:2005yk}. They may originate from decoherence, 
neutrino decay, or other physics mechanisms. In this section, we address this identification
in somewhat more detail in a qualitative manner. A relatively new ingredient for this
identification is the use of the ``Silver'' ($\nu_e \rightarrow \nu_\tau$) channel at a neutrino factory~\cite{Donini:2002rm,Autiero:2003fu}. It has been noticed \cite{Campanelli:2002cc,Kitazawa:2006iq,Friedland:2006pi} that the Silver channel probability can be greatly enhanced for non-standard Hamiltonian effects. This corresponds to what we have found in \Sec~\ref{sec:exp}, \ie, the Silver channel, which is similar to the ``Golden'' ($\nu_e \rightarrow \nu_\mu$) channel when there are no non-standard effects, behaves as our two-flavor limit in \Sec~\ref{sec:exp} for large energies.

\begin{figure}
\begin{center}
\includegraphics[width=14cm]{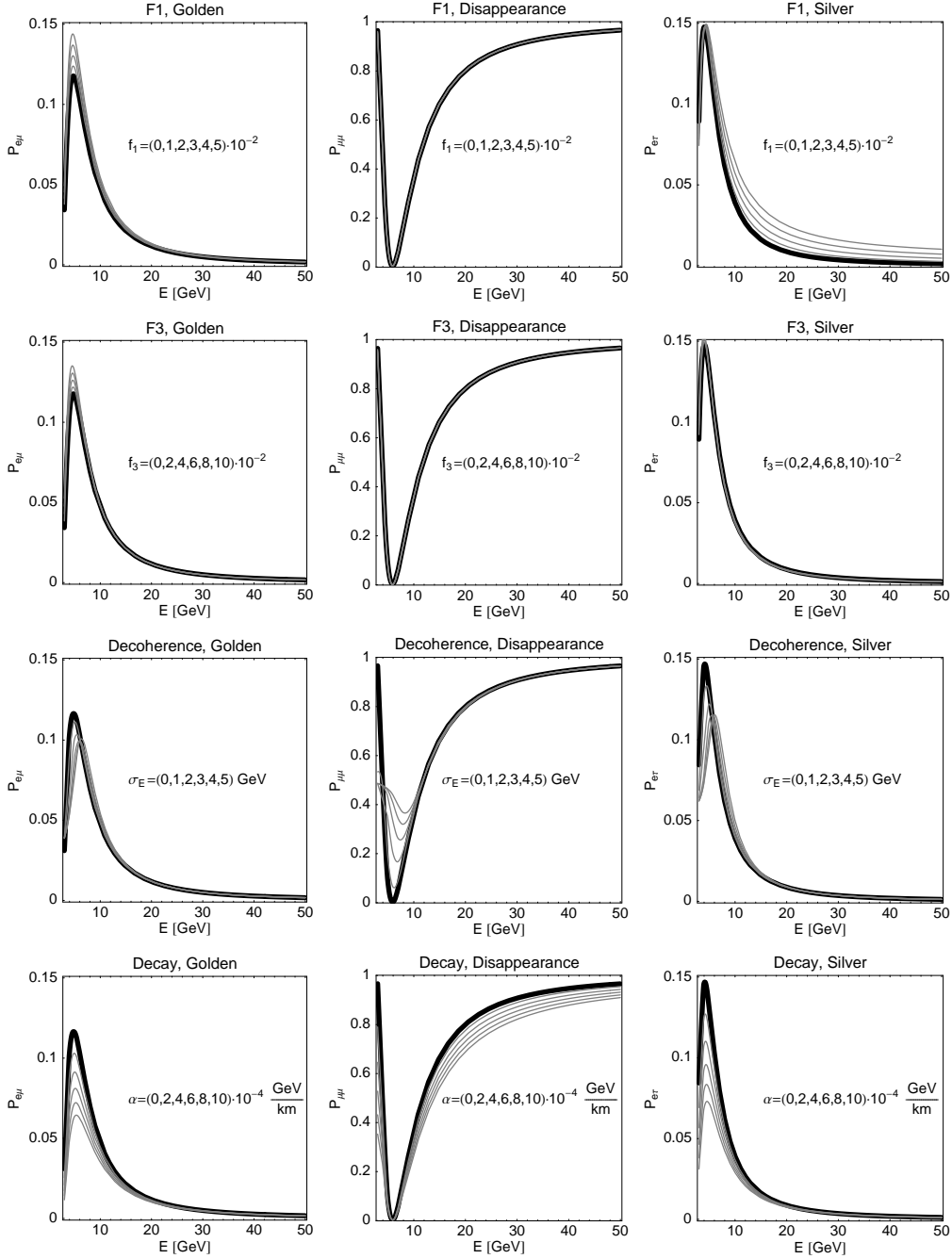}
\caption{The impact of four different non-standard effects (in rows) on three different oscillation channels (in columns): Golden $\nu_e \rightarrow \nu_\mu$, Disappearance $\nu_\mu \rightarrow \nu_\mu$, and Silver $\nu_e \rightarrow \nu_\tau$. The different rows correspond to $F_1$ (flavor-changing without \CP{} violation, Hamiltonian-level), $F_3$ (flavor-conserving, Hamiltonian-level), decoherence, and neutrino decay. The different model parameters for the different curves are given in the individual plots, where the thick curves correspond to the standard neutrino oscillation scenario. }
\label{fig:effectspanel}
\end{center}
\end{figure}

In \figu{effectspanel}, we show the impact of different types of effects on the neutrino oscillation probabilities in the Golden channel $P_{e\mu}$, the disappearance channel $P_{\mu\mu}$\footnote{The probability $P_{\mu\mu}$ is actually the $\nu_\mu$ survival probability. However, this is the relevant probability when searching for $\nu_\mu$ disappearance rather than the disappearance probability $1-P_{\mu\mu}$.}, and the Silver channel $P_{e\tau}$ (shown in columns) at a possible future neutrino factory (relevant energy range shown). The different rows correspond to scenarios with $F_1$ (flavor-changing without \CP{} violation, Hamiltonian-level), $F_3$ (flavor-conserving, Hamiltonian-level), decoherence, and neutrino decay, respectively. The different model parameters for the different curves are given in the individual plots, where the thick curves correspond to the standard neutrino oscillation scenario. For a description of the decoherence and decay models, see \Ref~\cite{Blennow:2005yk}. In short, the decoherence model corresponds to standard wave packet decoherence, whereas the decay model assumes equal decay rates for all mass eigenstates (which can, for instance, be motivated by a degenerate mass spectrum).
Note that this figure is shown for neutrinos only and that the comparison between the neutrino and antineutrino behavior can provide information on the underlying physics as well. However, this behavior is model-dependent.

Figure \ref{fig:effectspanel} is very useful to study the characteristics of different effects and to illustrate how the information from different neutrino oscillation channels can be used to disentangle them. First, it is important to note that it is difficult to construct a damping effect without large impact on the disappearance channel. Since the event rates in this channel are very high, it is probably the first place to look for non-standard physics. In addition, damping effects tend to suppress the Golden and Silver channel probabilities around the oscillation peak, which can, depending on the model, be very different for Hamiltonian-level effects. However, as it can be read off from \figu{effectspanel}, Hamiltonian-level effects produce, similar to $\stheta$, the largest effect in the appearance channels.\footnote{The weak influence of the non-standard Hamiltonian level effects on the disappearance probability $1-P_{\mu\mu}$ is purely due to the fact that \equ{numham} has been assumed for the non-standard Hamiltonian, where we have assumed the $\nu_\mu$ states to be unaffected. However, as mentioned earlier, there are also stronger bounds on any NSI involving $\nu_\mu$.} In particular, the flavor-conserving effect $F_1$ may enhance the silver probability as demonstrated in the two-flavor limit in \Sec~\ref{sec:exp}. Comparing all panels in \figu{effectspanel}, we expect that the combination of all channels serves as a good model discriminator because each of the shown models has a unique signature if all these channels are combined. For example, we have tested that adding a $5 \, \mathrm{kt}$ OPERA-like emulsion cloud chamber for the silver channel at the same baseline as the golden channel improves the $F_1$ discovery reach by about 60~\% (for simulation details, see \Ref~\cite{Huber:2006wb}) because of the silver channel signature at large energies. Therefore, we believe that this combination of different channels in combination with precise oscillation parameter measurements and spectral signatures can reveal non-standard physics.

\section{Summary and conclusions}
\label{sec:S&C}

For future long-baseline neutrino oscillation precision measurements, such as neutrino
factories, it will be an important question how to identify a non-standard effect.
While it is very likely that many theoretical models will fit such a ``new physics''
discovery, the classification of models corresponding to this discovery from a 
phenomenological point of view will be very important for the planning of the following
generation of experiments. Hence, there is strong interest in a top-down approach
to non-standard physics tests, since the impact of future measurements on theory
has to be assessed to promote the experiment. So far, mainly the bottom-up approach
has been used, which is testing specific models in an experiment. Therefore, it has
been one of the main goals of this work to demonstrate the identification and
separation of individual phenomenological classes by generic arguments.

In summary, we have studied non-standard effects on neutrino
oscillations on the Hamiltonian level. We have parameterized these effects
in terms of the generators of the Lie algebra, and we have 
introduced them in flavor (such as coming from FCNC or NSI) and 
mass (such as coming from MVN) bases.  As a trivial fact, there is, 
in principle, no mathematical difference between these
effects if one allows the most general form in each basis. 
Given the detection of a general non-standard effect on Hamiltonian level, it is therefore not possible to classify it as a flavor or mass effect without further assumptions or knowledge 
and, from an empirical point of view, the classification is a matter of definition. Therefore, we have
defined ``pure'' effects as effects which are proportional to specific
individual generators. Those correspond to pure flavor/mass
conserving/violating effects, \ie, effects which affect particular
flavor or mass eigenstates. This definition makes sense if one assumes
that the underlying theoretical model causes one dominating non-standard
effect. It is then the simplicity of the form in
the respective basis which defines the effect to be of flavor or mass
type. Therefore, the concept of these pure effects allows the choice of the most ``natural'' 
class of models for further testing, which is most appealing from the physics point of view.

{}From the analytical point of view, we have studied the effects in
the two-flavor limit.  We have derived the modified mass squared
differences and mixing angles (parameter mappings) as well as the
modified resonance conditions including standard matter effects. In
addition, we have discussed the application of this two-flavor limit
to experiments, in particular, to the neutrino oscillation probability $P_{e \mu}$. This probability can be
described to a first approximation by a two-flavor limit for large
$\stheta$, where the $\stheta$-term dominates the \CP{} effects.  In
addition, non-standard effects in the 1-3-sector have so far very poor
limits (such as $\epsilon_{e\tau}$) and the driving parameter
$\stheta$ is unknown, which means that there is room for confusion
between $\theta_{13}$ and non-standard effects (see, \eg,
\Ref~\cite{Huber:2001de,Huber:2002bi}). We have found that there are several
generic features for different types of effects. While any flavor
violating pure effect can obviously change the transition
probabilities, it does not affect the resonance
condition/energy. However, a flavor conserving pure effect changes the
resonance condition similar to matter effects and is highly correlated
with the matter density. In addition, it can suppress the flavor
transition for large energies similar to matter effects -- even in
vacuum. Pure mass effects behave, in principle, as rotations of the
flavor effects by the mixing angles, \ie, a pure mass effect will be
observed as a linear combination of flavor effects. However, for a
pure mass conserving effect, these flavor effects combine with special
characteristics, since the mass effect is similar to an (energy
dependent) change of the vacuum mass squared difference, \ie, it
basically squeezes or stretches the oscillation pattern.  Since in
quantum field theory any non-standard effect may originate in the
coherent (Hamiltonian effect) or incoherent (``damping'' effect)
summation of amplitudes, we have compared the non-standard Hamiltonian
effects to the previously studied ``damping'' effects on probability
level. We have found that these two classes can be distinguished by
typical characteristics. Non-standard Hamiltonian effects shift the
oscillation pattern, while ``damping'' effects, in general, do
not. In principle, the different classes of non-standard Hamiltonian effects can
be identified by their modification of oscillation
amplitudes for large energies, the shift of the matter resonance, the
comparison of different $L/E$-ranges, \etc

We have also studied some aspects of the three-flavor generalization
of general non-standard Hamiltonian effects using perturbation theory
as well as numeric calculations. By assuming small non-standard
Hamiltonian effects, we have derived expressions for the effective
matrix elements using perturbation theory and observed how the confusion theorem between $\theta_{13}$ and non-standard effects described in \Ref~\cite{Huber:2002bi} arises at the
Hamiltonian level. Our numeric calculations show that non-standard
effects can alter the determination of $\theta_{13}$
significantly at higher energies, while still preserving a high
accuracy at lower energies (\cf, \fig~\ref{fig:threeeff}).

Eventually, we have demonstrated, at a numerical example for a neutrino
factory, that many of these features can be found in a realistic experimental
simulation using three flavors and specific spectral (energy) dependencies
for the non-standard effects. For example, while it is simple
to distinguish a flavor changing effect from flavor conserving or mass effects in general,
mass effects are hard to establish as long as the neutrino oscillation parameters are not
known from an independent source (such as with a different matter density
for MVN). 
In addition, we have compared the obtainable discovery reaches for $\epsilon_{e \tau}$ 
to the current limits, and we have found at least an order of magnitude improvement
for large $\stheta$ and $\deltacp=0$.
We have also compared the Hamiltonian level effects to damping effects and found that they can be distinguished by their specific alteration of the spectra in different neutrino oscillation channels.

Since the past has told us that neutrinos are good for surprises, the
high precision measurements at future neutrino oscillation experiments might as
well reveal a detection of ``new physics'' beyond the Standard Model
(extended to include massive neutrinos). Therefore, we conclude that one
should include general strategies to look for non-standard effects in
future neutrino oscillation experiments, where we have followed a
top-down approach: Instead of testing particular models
(bottom-up), we have assumed that some inconsistency will be found
first. Secondly, one may want to classify this inconsistency to be
either a Hamiltonian or a probability level (``damping'') effect. Finally,
individual models are identified which fit this effect. Since we do
not know exactly what we are looking for, such an approach might be a
clever search strategy, and it can be useful to promote an experiment
as a discriminator among different classes of theoretical models. 
Future studies should demonstrate how such an
approach can be most efficiently extended to three neutrino flavors,
which neutrino oscillation channels are most suitable, and what the
correlations with the existing fundamental neutrino oscillation
parameters imply.

\subsection*{Acknowledgments}

W.W. would like to thank the Theoretical Elementary Particle Physics
group at KTH for the warm hospitality during a research visit. In
addition, T.O. and W.W. would like to thank Manfred Lindner and his
group at TUM in Munich for the warm hospitality during their research
visits where parts of this paper were developed.

This work was supported by the Royal Swedish Academy of Sciences
(KVA), the Swedish Research Council (Vetenskapsr{\aa}det), Contract
Nos.~621-2001-1611, 621-2002-3577, the G{\"o}ran Gustafsson
Foundation, the Magnus Bergvall Foundation, the W.~M.~Keck Foundation,
and NSF grant PHY-0070928.

\appendix

\section{General formalism for the two-flavor scenario}
\label{app:twoflavor}

Any two-flavor Hamiltonian can be written in
flavor basis on the form
\begin{equation}
H = \boldsymbol H \cdot \boldsymbol \sigma,
\end{equation}
where $\boldsymbol H \in \mathbb R^3$ and $\boldsymbol \sigma =
(\sigma_1,\sigma_2,\sigma_3)$ is the vector of the three Pauli
matrices (\cf, the pictorial description of two-flavor neutrino
oscillations in \Ref~\cite{Kim:1994dy}). For a time-independent
Hamiltonian, the time evolution operator is given by
\begin{equation}
S(t) = \exp(-{\rm i}Ht).
\end{equation}
Using the relation $(\boldsymbol A \cdot \boldsymbol \sigma)^2 =
|\boldsymbol A|^2$, one obtains
\begin{equation}
S(t) = \boldsymbol 1_2 \cos(|\boldsymbol H|t)
-{\rm i} \frac{H}{|\boldsymbol H|} \sin(|\boldsymbol H|t),
\end{equation}
where $\boldsymbol 1_2$ is the $2\times 2$ unit matrix.
This gives the two-flavor neutrino oscillation probabilities of the form
\begin{align}
P_{\alpha\alpha} &= \tr[P_+ S(t)] =
1 - \sin^2(2\tilde\theta)\sin^2(kt), \\
P_{\alpha\beta} &= \tr[P_- S(t)] = \sin^2(2\tilde\theta)\sin^2(kt),
\end{align}
where
$$
P_\pm = \frac{1\pm \sigma_3}{2}, \quad
\sin^2(2\tilde\theta) = \frac{H_1^2 + H_2^2}{|\boldsymbol H|^2}, \quad
\mbox{and} \quad k = |\boldsymbol H| = \sqrt{\sum_{i=1}^3 H_i^2}.
$$
Here the $H_i$'s are the components of the Hamiltonian and
$\tilde\theta$ is the effective mixing angle.

In the standard two-flavor neutrino oscillation scenario, $H_1 =
\sin(2\theta)\Delta m^2/(4E)$, $H_2 = 0$, and $H_3 = V/2 -
\cos(2\theta)\Delta m^2/(4E)$. In general, the resonance condition,
\ie, the condition for maximal effective mixing, is $H_3 = 0$.  The
Hamiltonian is represented as a vector in $\mathbb R^3$, the third
direction being the ``flavor'' eigendirection. The mixing is given by
the angle between the Hamiltonian vector and the flavor
eigendirection. The mixing is maximal, {\it i.e.},
$\sin^2(2\tilde\theta) = 1$, when the Hamiltonian vector is orthogonal
to the flavor eigendirection, which, as expected, is equivalent to the
resonance condition.

The flavor and mass bases, and thus, the flavor and mass effects, are
intimately associated with each other. For the case of $n=2$, \ie, for
two neutrino flavors, any effective contribution to the Hamiltonian
can be written in either flavor or mass basis, \ie, as $H' = F_1 \rho_1
+ F_2 \rho_2 + F_3 \rho_3$ or $H' = M_1 \tau_1 + M_2 \tau_2 + M_3
\tau_3$. Since the effect must be the same regardless of
the basis it is expressed in, we obtain the relations
\begin{equation}
\left\{ \begin{array}{l} F_1 = M_1 \cos(2\theta) - M_3 \sin(2\theta)
  \\ F_2 = M_2 \\ F_3 = M_1 \sin(2\theta) + M_3 \cos(2\theta)
  \end{array} \right.
\label{equ:flavor_mass_transformation}
\end{equation}
from \equ{sigma_tau}, \ie, one obtains $F_1$ and $F_3$ by rotating
$M_1$ and $M_3$ by the angle $-2\theta$ as well as one has $F_2 =
M_2$. Thus, the transformation in \equ{flavor_mass_transformation}
relates flavor and mass effects and shows that they are linear
combinations of each other.

\section{Non-standard effects for large mixing}
\label{app:visualization}

\begin{figure}[t]
\begin{center}
\includegraphics[width=14cm,clip]{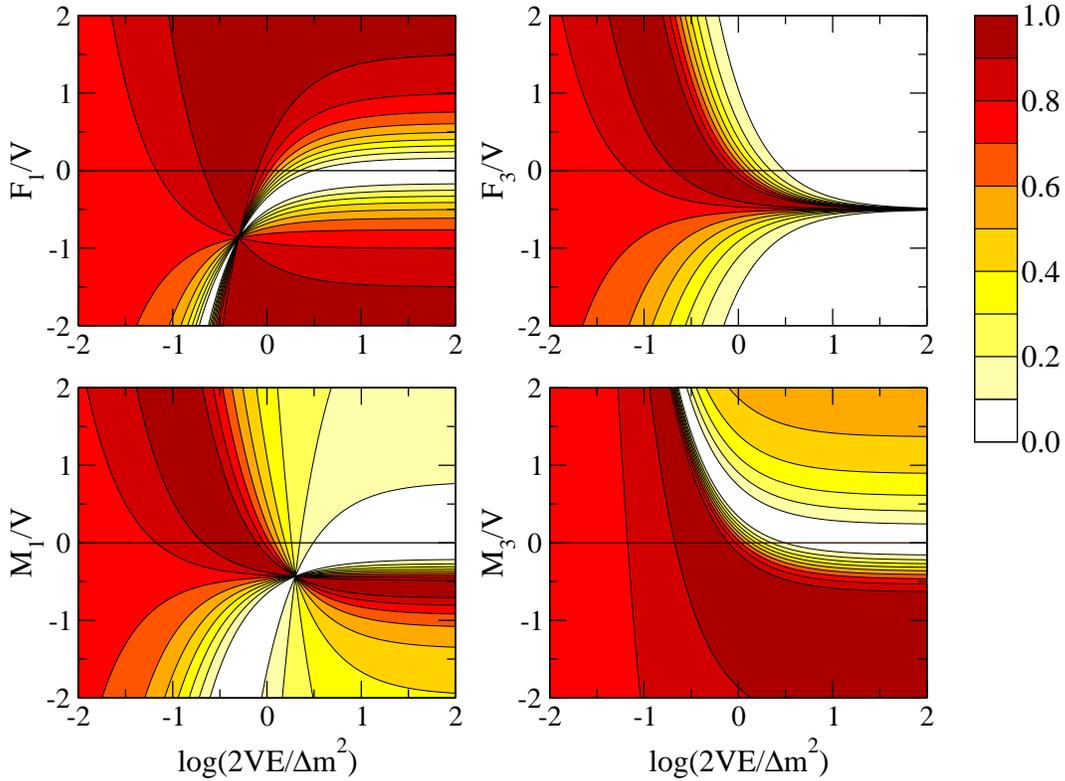}
\caption{The effective mixing $\sin^2(2\tilde\theta)$ as a function of
$2VE/\Delta m^2$ and the ratio between the pure flavor ($F_1$ and
$F_3$) or mass ($M_1$ and $M_3$) effects and the matter potential
$V$. The horizontal lines correspond to no non-standard effect and the
mixing along them is therefore the same in all panels. The vacuum
mixing is assumed to be $\theta = 30^\circ$, which is close to the
present value of the solar mixing angle $\theta_{12}$ (see, \eg,
\Ref~\cite{Maltoni:2004ei}). See the main text for a more detailed
discussion.}
\label{fig:mixingeffects}
\end{center}
\end{figure}

In this appendix, we concentrate on pure effects in the limit of large
mixing. When the mixing goes to maximal, we have $\cos (2 \theta)
\rightarrow 0$. This means that the resonance condition in
\equ{rescondflavor} cannot be fulfilled for $F_3=0$ (\ie, ``matter
resonance'') at reasonably large energies.\footnote{Note that, in this
appendix, we assume that matter effects determine the resonance energy
and the non-standard effects are sub-leading contributions, which may
shift the resonance energy. Thus, we refer to the ``matter resonance''
as the resonance condition in \equ{rescondflavor} for $F_3 = 0$.}
{}From \equ{thetaflavor}, we can easily observe that $F_1$ and $F_2$
will not modify $\sin^2 (2 \tilde{\theta})$ at all in the absence of
matter (and $F_3$) effects (for example, in $P_{\mu \mu}$ or a vacuum
probability). Independent of matter effects, $F_3$ can increase the
suppression of $\sin^2 (2 \tilde{\theta})$ for large energies. If the
resonance condition in \equ{rescondflavor} is fulfilled, then $\sin^2
(2 \tilde{\theta})$ will be independent of $F_1$ and $F_2$.  However,
in the presence of matter effects (such as for $P_{ee}$ in the limit
$\theta_{13} \rightarrow 0$), $F_1$ and $F_2$ can reduce the matter
effect suppression of $\sin^2 (2 \tilde{\theta})$ for large energies
(\cf, \Fig~\ref{fig:mixingeffects}), \ie, they can increase the
effective mixing. Eventually, it is obvious from
\eqs~(\ref{equ:dmflavor}) and (\ref{equ:xiflavor}) that the
oscillation frequency is always increased for positive $F_i/\Delta
m^2$'s.

It can also be interesting (and quite illuminating) to study how
different pure (flavor or mass) effects affect the effective neutrino
mixing and oscillations. In \fig~\ref{fig:mixingeffects}, we plot
the effective mixing resulting from ``pure'' flavor and mass effects.
{}From this figure, some features become quite apparent, \eg, some
generic features are the shift in the resonance energy for $F_3$,
$M_1$, and $M_3$, the non-zero high-energy mixing for all effects but
the flavor conserving effect $F_3$, and the appearance of an anti-resonance
-- where $\sin^2(2\tilde\theta)$ goes to zero for some finite energy
-- for $F_1$, $M_1$, and $M_3$.

The shift in the resonance energy is simply due to the shift in the
$H_3$-component of the total effective Hamiltonian in flavor basis (as
was mentioned earlier, the resonance condition is $H_3 = 0$). The fact
that there is no shift of the resonance condition for $F_1$ and $F_2$
was also discussed earlier.

The reason why the effective high-energy mixing generally turns out to
be non-zero is also quite easy to realize. At high energies, the
effective matter potential, which is diagonal in flavor basis,
dominates over the vacuum Hamiltonian. As a result, the effective
mixing is usually zero at high energies. However, if there is a
non-standard effect with a corresponding effective addition to the
Hamiltonian which is non-diagonal and is either constant or increasing
with energy, then the effective mixing at high energies will be fully
determined by the ratio of the non-standard effect and the matter
potential.

The anti-resonance appears when $H_1 = H_2 = 0$ in flavor
basis. Since $H_2 = 0$ in the standard neutrino oscillation scenario,
it is apparent that this anti-resonance will occur for some value of
$F_1$. In addition, since $M_1$ and $M_3$ are linear combinations of
$F_1$ and $F_3$, the anti-resonance will also appear for $M_1$ and
$M_3$ effects, as can be seen from the plots in
\Fig~\ref{fig:mixingeffects}.

There are also some interesting features that are specific for
different effects. First, for $F_1$ effects, we note that the
resonance condition is unchanged and that the mixing is constant as a
function of energy for $F_1/V = -\tan(2\theta)/2$ (the reason for this
is that the sum of the non-standard Hamiltonian and the matter
potential is proportional to the vacuum Hamiltonian).  Then, for
flavor conserving $F_3$ effects, we note that these correspond to
changes in the effective matter potential. For $F_3/V < -1/2$, we
obtain an effective matter potential which is negative, resulting in a
disappearance of the resonance.  Next, for $M_1$ effects, the mixing
is constant when $M_1/V = -\sin(2\theta)/2$, in analogy with the $F_1$
effects (again, the reason is that the sum of the non-standard
Hamiltonian and the matter potential is proportional to the vacuum
Hamiltonian). Also in analogy with the $F_1$ effects is that there is
a value of $2VE/\Delta m^2$, where the mixing does not depend on the
$M_1/V$ ratio. However, in the case of $M_1$ effects, this is not the
resonance mixing, but rather a mixing of $\sin^2(2\tilde\theta) =
\cos^2(2\theta)$, which appears at $2VE/\Delta m^2 =
1/\cos(2\theta)$. Finally, for mass conserving $M_3$ effects, we note
that the resonance disappears when $2\cos(2\theta)M_3/V < -1$.

The reason why the equivalent $F_2$ and $M_2$ effects are not included
is that these effects always lead to an increase in the effective
mixing angle for all energies, and thus, those plots do not show as
many interesting features as the plots included. In addition, we note
that if the non-standard effects are energy dependent, then the effective
mixing will be given by the mixing along some non-constant function of
$2VE/\Delta m^2$ in \fig~\ref{fig:mixingeffects}.

\section{Two-flavor limits of three flavor scenarios}
\label{app:Pemu}

In this appendix, we discuss subtleties with the definition of the effective two-flavor scenarios introduced in \Sec~\ref{sec:exp}.
Remember that the effective two-flavor neutrino oscillation
scenarios should be defined in terms of the effective two-flavor
sector in question. For example, in the limit when $\Delta m_{21}^2
\rightarrow 0$, the effective two-flavor sector is spanned by $\nu_e$
and $\nu_a = - s_{23} \nu_\mu + c_{23} \nu_\tau$. Thus, the limit can
be considered as an exact pure two-flavor scenario only if the non-standard
effects preserve the two-flavor limit (\ie, no off-diagonal terms
mixing $\nu_e$ and $\nu_a$ with the remaining neutrino state $\nu_b =
c_{23}\nu_\mu + s_{23}\nu_\tau$). If the non-standard addition to the
Hamiltonian is given by
\begin{equation}
	H'_{\alpha\beta} = \eps_{\alpha\beta} V,
\end{equation}
then the corresponding addition in the basis spanned by $\{\nu_e, \nu_b,
\nu_a\}$ is
\begin{equation}
	H' = V\left(\begin{array}{ccc}
	\eps_{ee} & c_{23}\eps_{e\mu}-s_{23}\eps_{e\tau} & c_{23}\eps_{e\tau} + s_{23}\eps_{e\mu} \\
	c_{23}\eps_{e\mu}^* - s_{23}\eps_{e\tau}^* & A & B \\
	c_{23}\eps_{e\tau}^* + s_{23}\eps_{e\mu}^* & B^* & C
	\end{array}\right),
\end{equation}
where
\begin{eqnarray*}
	A &=& c_{23}^2\eps_{\mu\mu}+s_{23}^2\eps_{\tau\tau}-s_{23}c_{23}(\eps_{\mu\tau}+\eps_{\mu\tau}^*),
	\\
	B &=&
	c_{23}\eps_{\mu\tau}-s_{23}^2\eps_{\mu\tau}^*+s_{23}c_{23}(\eps_{\mu\mu}-\eps_{\tau\tau}),
	\\
	C &=&
	s_{23}^2\eps_{\mu\mu}+c_{23}^2\eps_{\tau\tau}+s_{23}c_{23}(\eps_{\mu\tau}+\eps_{\mu\tau}^*).
\end{eqnarray*}
{}From this relation, we deduce that the limit will be a pure two-flavor
case if $\eps_{\alpha\beta} = 0$ for all non-standard effects which do
not involve $\nu_e$ and $c_{23}\eps_{e\mu}-s_{23}\eps_{e\tau} = 0$
(which could be implemented by, \eg, $\theta_{23} = 45^\circ$ and
$\eps_{e\mu} = \eps_{e\tau}$). In general, some of the conclusions for
the two-flavor case will therefore not apply to three flavors. We
have, in the numerical example in \Sec~\ref{sec:numex}, demonstrated
which of the conclusions that do hold. The case when $\theta_{13} \rightarrow 0$ is similar to the case described above, with the exception that the effective two-flavor sector is now spanned by $\nu_e$ and $\nu_b$ instead of $\nu_e$ and $\nu_a$. For the limit $\theta_{13} \to 0$ and $\Delta m_{32}^2 \rightarrow 0$, there is no subtlety and the two-flavor sector is spanned by $\nu_\mu$ and $\nu_\tau$.

\end{document}